\documentclass{article}
\usepackage{spconf,amsmath,graphicx}
\usepackage{cite}
\usepackage{algorithmic}
\usepackage{algorithm}
\usepackage{multirow}
\usepackage{subfigure}
\usepackage{rotating}
\usepackage{array}
\renewcommand{\vec}[1]{\boldsymbol{#1}} 

\title{Identifying Rumor Sources Using Dominant Eigenvalue of Nonbacktracking Matrix}
\name{Jiachun Pan, Wenyi Zhang\thanks{This work was supported in part by the National Natural Science Foundation of China under Grant 61722114.}}
\address{University of Science and Technology of China, Hefei, China\\Emails: pjc1993@mail.ustc.edu.cn, wenyizha@ustc.edu.cn}
\begin{document}
\ninept
\maketitle
\begin{abstract}
We consider the problem of identifying rumor sources in a network, in which rumor spreading obeys a time-slotted susceptible-infected model. Unlike existing approaches, our proposed algorithm identifies as sources those nodes, which when set as sources, result in the smallest dominant eigenvalue of the corresponding reduced nonbacktracking matrix deduced from message passing equations. We also propose a reduced-complexity algorithm derived from the previous algorithm through a perturbation approximation. Numerical experiments on synthesized and real-world networks suggest that these proposed algorithms generally have higher accuracy compared with representative existing algorithms.
\end{abstract}
\section{Introduction}
\label{sec:intro}

Nowadays, facilitated by the development of Internet and smart devices, online social networks such as Twitter, Facebook, and Weibo have become important enablers and primary conduits of rumor-like information. It is thus of considerable interest and importance to accurately identify rumor sources in networks.

Several works have investigated the problem of identifying rumor sources given a snapshot observation of the infected nodes in a network. For the basic case of a single rumor source, a network centrality metric called rumor center was initially developed in \cite{reference6}, which turns out to be the maximum likelihood estimate of the rumor source for regular tree networks under the susceptible-infected (SI) rumor spreading model. This has inspired a large number of works. For example, in \cite{reference7} \cite{reference12} another network centrality metric called Jordan center was used to detect a single rumor source under the susceptible-infected-recovered (SIR) rumor spreading model; in \cite{reference9} the problem of rumor source identification was solved using Monte Carlo estimators, for an arbitrary network structure; in \cite{reference8} the case where multiple snapshot observations are available was studied and a joint estimator was developed.

An important extension of the basic case is the case where multiple rumor sources exist. In \cite{reference14} a method based on rumor centrality was developed, with computational complexity $O(N^{|S|})$, where $N$ is the number of infected nodes and $|S|$ is the number of rumor sources. In \cite{reference15} a method based on Jordan Center was developed under the SIR model, and it was proved that for regular trees the identified sources are within a constant distance from the actual sources with a high probability. In \cite{reference16} a K-center method was developed, which first transforms the network into a distance network using an effective distance metric, then adaptively partitions the distance network and finally performs source identification. The method has a complexity of $O(MN\log\alpha)$, where $\alpha$ is a slowly growing inverse-Ackermann function related to the number of nodes $N$ and the number of edges $M$.

We note two facts: first, all the existing methods for both the single source and the multiple source cases are not optimal when networks contain loops; second, all the existing methods for the multiple source case need to partition the infected network and then identify a source in each partitioned part, regardless of the reality that the infected nodes from different sources may substantially overlap.

In this paper, we investigate the problem of identifying multiple rumor sources in a general network which may be highly loopy, and our main contribution is the proposal of a novel heuristic method based on the dominant eigenvalue of a matrix obtained from the Hashimoto or nonbacktracking matrix \cite{reference10} of the infected network. We develop the method via an approximate analysis of message passing equations of the rumor spreading process, combined with some empirical observations. Unlike existing methods, our method neither needs to convert a loopy network into a tree, nor needs to partition the infected network into non-overlapping parts. Numerical experiments on synthesized and real-world networks suggest that our method generally has higher accuracy compared with representative existing algorithms, especially for highly loopy networks.

The rest of the paper is organized as follows. Section \ref{sec:problem} describes the problem formulation, message passing equations, and the proposed approach. Section \ref{sec:algorithm} presents the algorithms. Section \ref{sec:simu} shows the numerical results. Section \ref{sec:conclusion} concludes the paper.

\section{Problem Formulation, Message Passing Equations, and Proposed Approach}
\label{sec:problem}

In this section, we describe the problem setup, develop the message passing equations, and motivate our approach.

\subsection{Rumor spreading model}
\label{subsec:SI}

We assume that rumors spread in an undirected graph $G = (V, E)$, where $V$ is the set of nodes and $E$ is the set of edges. We adopt a time-slotted susceptible-infected (SI) model. In this model, each node $v \in V$ has two possible states: susceptible and infected. A source node is infected from the beginning; a non-source node is said to be susceptible if it has not received the rumor, and infected if it has received the rumor from one of its neighbors. We assume that there are $|S|$ rumor sources, and they start to initialize their rumor spreadings at the same time, termed time \emph{zero}. At each time step, an infected node infects a neighboring susceptible node with probability $p \ll 1$, and such infections of different neighboring susceptible nodes are mutually independent. As time grows without bound, eventually all connected nodes will be infected for any non-zero value of $p$. Given a snapshot observation of the infected nodes, we want to identify those rumor sources.

\subsection{Message passing equations}
\label{subsec:MP}

Denote by $P_i^{(t)}$ the probability that node $i$ has \emph{not} been infected by time $t$, and by $v_{i\rightarrow j}^{(t)}$ the probability that node $i$ has \emph{not} passed the rumor to a neighboring node $j$ by time $t$. Note that by assumption the rumor spreading starts at time $t = 0$. Assign to each node an indicator $n_i$ to indicate whether node $i$ is a source: $n_i = 0$ if node $i$ is a source, and $n_i = 1$ otherwise.

Denote by $\partial i$ the set of neighbors of node $i$, and $\partial i\setminus j$ the set of neighbors of node $i$ excluding node $j$. For a general network with loops, the probabilities $\left\{v_{k\rightarrow i}^{(t)} \big| k\in \partial i \setminus j\right\}$ are to some extent correlated and this fact makes a rigorous analysis intractable. Therefore, in subsequent development we make a key approximation that these probabilities are mutually independent; see, e.g., \cite{reference5} for a similar treatment. Hence we can deduce the following message passing equations:
\begin{equation}\label{eqn:message-passing}
v_{i\rightarrow j}^{(t)} = 1 - \sum_{\tau=1}^{t}(1-p)^{\tau-1} p\left[1-n_i\prod_{k\in \partial i \setminus j}{v_{k\rightarrow i}^{(t-\tau)}}\right].
\end{equation}
To see the validity of (\ref{eqn:message-passing}), note that if node $i$ is a source, $n_i = 0$, the equation (\ref{eqn:message-passing}) is simply the probability that node $i$ has not infected node $j$ throughout $t$ time steps, with the infection probability at each time being a Bernoulli distribution; otherwise, if $n_i = 1$, the term in the box bracket of (\ref{eqn:message-passing}) is the probability that node $i$ has been infected by time $t - \tau$ (under the approximation of mutual independence), the term $(1-p)^{\tau-1} p$ is the probability that node $i$ then takes $\tau$ time steps to infect node $j$, and (\ref{eqn:message-passing}) then follows from the law of total probability.

We can also deduce the probabilities $P_i^{(t)}$ as:
\begin{equation}
P_i^{(t)} = n_i \prod_{j\in\partial i}v_{j\rightarrow i}^{(t)}.
\end{equation}

Taking the limit $t\rightarrow\infty$ in (1), and noting that $(1-p)^{\tau-1} p \rightarrow 0$ as $\tau \rightarrow \infty$, we get
\begin{equation}
v_{i\rightarrow j}^{(\infty)} = n_i\prod_{k\in \partial i \setminus j}{v_{k\rightarrow i}^{(\infty)}}.
\end{equation}
This confirms that once a node is infected, its neighbors will eventually be infected as time grows without bound.

\subsection{Linear approximation}
\label{subsec:linear-approximation}

A snapshot observation of the infected nodes is a graph with $N$ nodes and $M$ edges. For a given snapshot observation, the approximate message passing equations in (1) constitute a collection of nonlinear equations of $v_{\rightarrow}^{(\tau)} = \left(\cdots, v_{i\rightarrow j}^{(\tau)}, \cdots\right)$ for all the $2M$ directed links $i \rightarrow j$ and all time steps $\tau \in \{0, 1, \ldots, t\}$, which can be collectively written as a discrete-time dynamical system with initial condition $v_{\rightarrow}^{(0)} = \vec{e}$:
\begin{equation}\label{eqn:evolution-nonlinear}
v_{\rightarrow}^{(t)}=\vec{e}-\sum_{\tau=1}^t(1-p)^{\tau-1}p[\vec{e}-\vec{f}\left(v_{\rightarrow}^{(t-\tau)}, n_\rightarrow\right)],
\end{equation}
where $\vec{e}$ is a $1\times2M$ all-one vector, $n_{\rightarrow}=(\cdots,n_{i\rightarrow j},\cdots)$, in which $n_{i\rightarrow j}=n_i$ and $\vec{f}=(\cdots,f_{i\rightarrow j},\cdots)$ in which $f_{i\rightarrow j}$ is the nonlinear function of $v_{\rightarrow}$ for $i\rightarrow j$ given in (1).

To obtain some insights, we linearly approximate $f_{i\rightarrow j}(v_{\rightarrow},n_{\rightarrow})$ at $v_{\rightarrow}=\vec{e}$:
\begin{equation}
f_{i\rightarrow j}(v_{\rightarrow },n_{\rightarrow})=f_{i\rightarrow j}(\vec{e},n_{\rightarrow})+f^\prime_{i\rightarrow j}(\vec{e},n_{\rightarrow})(v_{\rightarrow}-\vec{e}),
\end{equation}
where $f^\prime_{i\rightarrow j}(v_{\rightarrow},n_{\rightarrow})=\left(\cdots,\frac{\partial f_{i\rightarrow j}}{\partial v_{k\rightarrow l}},\cdots\right)$. By (1), we have that for $l \neq i$, $\frac{\partial f_{i\rightarrow j}}{\partial v_{k\rightarrow l}}\bigg|_{v_{\rightarrow}=\vec{e}}=0$, and for $l=i$ and $k\neq j$, $\frac{\partial f_{i\rightarrow j}}{\partial v_{k\rightarrow l}}\bigg|_{v_{\rightarrow}=\vec{e}}=n_i$. So we get the linear approximation of $v_{i\rightarrow j}^{(t)}$ as
\begin{equation}
v_{i\rightarrow j}^{(t)}=1-\sum_{\tau=1}^t(1-p)^{\tau-1}p\left[1-n_i-n_i \sum_{k \in \partial i \setminus j}\left(v_{k\rightarrow i}^{(t-\tau)}-1\right)\right],
\end{equation}
which can be collectively written in matrix form as
\begin{equation}\label{eqn:message-passing-linear}
v_{\rightarrow}^{(t)}=\vec{e}-\sum_{\tau=1}^t(1-p)^{\tau-1}p\left[\vec{e}-n_{\rightarrow}+(\vec{e}-v_{\rightarrow}^{(t-\tau)})\textit{R}\right],
\end{equation}
where $\textit{R}$ is a $2M \times 2M$ matrix:
\begin{equation}
\textit{R}_{k\rightarrow l,i\rightarrow j}=n_i\textit{B}_{k\rightarrow l,i\rightarrow j},\;\textit{B}_{k\rightarrow l, i\rightarrow j}=
\begin{cases}
1&\text{if $l=i$ and $j\neq k$}\\
0&\text{otherwise}.
\end{cases}
\end{equation}
The matrix $\textit{B}$ is known as the Hashimoto or nonbacktracking matrix of a graph \cite{reference10}, which is also closely associated with the spreading capability \cite{reference4}. We thus call $\textit{R}$ a reduced nonbacktracking matrix since it is obtained from $\textit{B}$ by setting entries to zero corresponding to $n_i = 0$.

We can further rewrite (\ref{eqn:message-passing-linear}) in a recursive form so that $v_{\rightarrow}^{(t+1)}$ is only related to $v_{\rightarrow}^{(t)}$:
\begin{equation}\label{eqn:message-passing-linear-recursive}
u_{\rightarrow}^{(t+1)} = p(\vec{e}- n_{\rightarrow})+u_{\rightarrow}^{(t)}[p\textit{I}+(1-p)\textit{R}],
\end{equation}
where $\textit{I}$ is the identity matrix, $u_{\rightarrow}^{(t)} = \vec{e} - v_{\rightarrow}^{(t)}$ in which $u_{i\rightarrow j}^{(t)}=1-v_{i\rightarrow j}^{(t)}$ is the probability that node $i$ has passed the rumor to a neighboring node $j$ by time $t$.

\subsection{Proposed approach}
\label{subsec:approach}

\begin{figure}[ht]
\centering
\centerline{\includegraphics[width=0.9\linewidth]{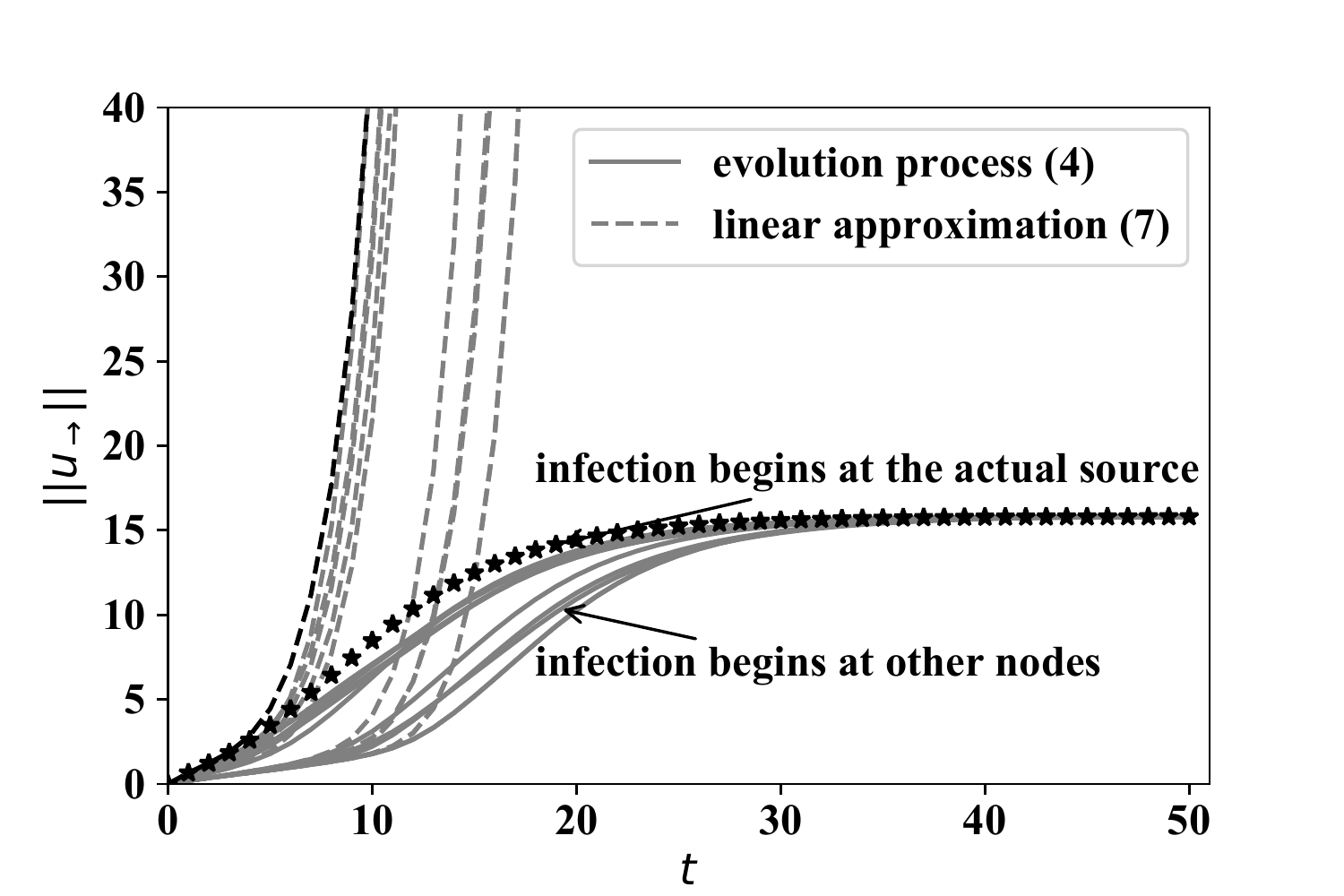}}
%
\caption{Evolution processes of $\|u_{\rightarrow}\|$ in a small-world network.}
\label{fig1}
\end{figure}

The task of identifying rumor sources is to determine a $\{0, 1\}$-vector $\vec{n}$ subject to $\sum_{i = 1}^N n_i = N - |S|$. Different choices of $\vec{n}$ lead to different evolution processes of $v_{\rightarrow}^{(t)}$ (or $u_{\rightarrow}^{(t)}$ equivalently). In our study, we have numerically computed such evolution processes, according to both (\ref{eqn:evolution-nonlinear}) and its linear approximation (\ref{eqn:message-passing-linear}), for different types of networks. Fig. \ref{fig1} displays a representative scenario, where for a snapshot observation of the infected graph in a small-world network containing a single rumor source (i.e., $|S| = 1$),\footnote{Here the number of infected nodes is 400 and $p$ is 0.1.} we compute and draw $\|u_{\rightarrow}^{(t)}\|$ under different choices of $\vec{n}$. The dark star curve is the evolution process of $\|u_{\rightarrow}^{(t)}\|$ when $\vec{n}$ coincides with the actual rumor source, the grey solid curves are the evolution processes of $\|u_{\rightarrow}^{(t)}\|$ when $\vec{n}$ randomly indicates a rumor source, and the dashed curves are the linear approximations of $u_{\rightarrow}^{(t)}$.

From Fig. \ref{fig1}, we have the following empirical observations:
\begin{itemize}
\item The evolution processes of $\|u_{\rightarrow}^{(t)}\|$ computed according to (\ref{eqn:evolution-nonlinear}) eventually approach the stable state where every connected node is infected with probability one, and when $\vec{n}$ coincides with the actual rumor source, the evolution process approaches this stable state the most quickly.
\item The linear approximation of $u_{\rightarrow}^{(t)}$ computed according to (\ref{eqn:message-passing-linear}) is accurate for small values of $t$, so that when $\vec{n}$ coincides with the actual rumor source, the linearly approximated evolution process of $\|u_{\rightarrow}^{(t)}\|$ grows the most quickly.
\end{itemize}

Iteratively applying (\ref{eqn:message-passing-linear-recursive}) leads to:
\begin{equation}\label{eqn:iteration}
\begin{aligned}
u_{\rightarrow}^{(t)}=&p(\vec{e}-n_{\rightarrow})\left(\textit{I}+[(1-p)\textit{I}+p\textit{R}]+[(1-p)\textit{I}+p\textit{R}]^2+\right.\\
&...+\left.[(1-p)\textit{I}+p\textit{R}]^{t-1}\right).
\end{aligned}
\end{equation}
So in the linear approximation, the growth rate of $\|u_{\rightarrow}^{(t)}\|$ is determined by $(\vec{e}-n_\rightarrow)[(1-p)\textit{I}+p\textit{R}]^t \approx (\vec{e}-n_\rightarrow)[(1-p)\textit{I}+p\textit{B}]^t$ since $p \ll 1$ and $\textit{R}$ differs from $\textit{B}$ only at the few entries indicated by $n_\rightarrow$. In light of the two empirical observations, now the task is to choose $n_\rightarrow$ such that the non-zero entries of $(\vec{e}-n_{\rightarrow})$ select the rows of $[(1-p)\textit{I}+p\textit{B}]^t$ whose sum vector yields the largest norm. With a binomial expansion, this norm is determined by the numbers of nonbacktracking paths of lengths up to $t$ starting from edges like $s \rightarrow i$ where $s$ is a source node indicated by $\vec{n}$ and $i$ is one of its neighbors. On the other hand, note that choosing $n_\rightarrow$ also determines the reduced nonbacktracking matrix $\textit{R}$, and that when the resulting $\textit{R}$ has the smallest dominant eigenvalue, the corresponding sources have the largest number of nonbacktracking paths passing them \cite{reference4}. This observation provides a reasonable heuristic for the choice of $n_\rightarrow$,\footnote{This heuristic has been verified with extensive numerical experiments in our study.} and motivates us to propose the following minimax criterion:
\begin{equation}\label{eqn:minimax}
\begin{aligned}
\min_{\vec{n}} \max \lambda(\textit{R}).
\end{aligned}
\end{equation}

\section{Algorithms}
\label{sec:algorithm}

In this section, we describe algorithms to identify rumor sources based on the reasoning in the previous section.

\subsection{Multiple source identification (MSI) algorithm}
\label{subsec:original}

According to (\ref{eqn:minimax}), we need to find the nodes, which when set as sources, result in the minimum dominant eigenvalue of the corresponding reduced nonbacktracking matrix. We can use the power iteration to compute the dominant eigenvalue in a time that scales as $O(M)$,\footnote{In our simulations in Section \ref{sec:simu} the number of iterations is fixed as $20$.} and repeat this for each of the $\binom{N}{|S|}$ configurations of sources. Thus, the complexity for the MSI algorithm is $O(M N^{|S|})$. The procedure of the MSI algorithm is in Table \textbf{Algorithm 1}.

\begin{algorithm}[htb]
\caption{Multiple Source Identification (MSI)}
\begin{algorithmic}[1]\label{algorithm:msi}
\REQUIRE ~~\\
Nonbacktracking matrix $\textit{B}$ of an infected graph of $N$ nodes and $M$ edges;\\
Number of rumor sources $|S|$;
\ENSURE ~~\\
Identified rumor sources $\hat{S}$;
\FOR{$i = 1, \ldots, \binom{N}{|S|}$}
\STATE{Enumerate a set of potential source nodes as $\hat{S}_i= \{i_1,i_2,...,i_{|S|}\}$;}
\STATE{Form $\textit{R}_i$ by setting the entries in $\textit{B}$ to zero corresponding to $n_{i_j} = 0$, $i_j\in\hat{S}_i$;}
\STATE{Use the power iterative to compute the dominant eigenvalue $\lambda_{i, \max}$ of $\textit{R}_i$;}
\ENDFOR
\STATE{Declare the set $\hat{S}_i$ with the minimal $\lambda_{i, \max}$ as rumor sources $\hat{S}$.}

\end{algorithmic}
\end{algorithm}

\subsection{Perturbation-based multiple source identification (PMSI) algorithm}
\label{sub:perturbation}
Now we derive an approximation of $\Delta{\lambda}=\lambda_{\max}(\textit{B})-\lambda_{\max}(\textit{R})$ by applying a method similar to that in \cite{reference20}. Denote $\textit{R}$ by $\textit{B}-\Delta\textit{B}$, the dominant eigenvalue of $\textit{R}$ by $\lambda_{\max}-\Delta\lambda$ and its corresponding right eigenvector by $u -\Delta u$. We have
\begin{equation}\label{eqn:difference-form}
(\textit{B}-\Delta{\textit{B}})(u-\Delta{u})=(\lambda_{\max}-\Delta{\lambda})(u-\Delta{u}).
\end{equation}
where $\lambda_{\max}$ is the dominant eigenvalue of $\textit{B}$.
Then we multiply both sides of (\ref{eqn:difference-form}) by the left eigenvector $v^T$ and get
\begin{equation}
\Delta{\lambda}=\frac{v^T\Delta{\textit{B}}u-v^T\Delta{\textit{B}}\Delta{u}}{v^Tu-v^T\Delta{u}}.
\end{equation}

We apply a perturbation analysis on $\Delta{u}$. When we set entries in $\textit{B}$ to zero according to the source set $S$, the entries $i\rightarrow s$ ($s\in S$, $i\in \partial{s}$) in $u$ will be zero, and other entries will be perturbed slightly. So we write $\Delta{u}=u_{\rightarrow s}-\delta u$, where $u_{\rightarrow s}$ is a vector in which we only keep the entries $i\rightarrow s$ ($s\in S$, $i\in \partial{s}$) of $u$ and set others to zero, and $\delta u$ is small. So by neglecting second order terms $u^T\Delta{\textit{B}}\delta u$ and $\Delta{\lambda} u^T\delta u$, we obtain
\begin{equation}
\Delta{\lambda} \approx\frac{v^T\Delta{\textit{B}}u-v^T\Delta{\textit{B}}u_{\rightarrow s}}{v^Tu-v^Tu_{\rightarrow s}}.
\end{equation}
According to the definition of $\Delta{\textit{B}}$, we obtain
\begin{equation}\label{eqn:pmsi}
\Delta{\lambda}\approx\frac{\sum_{i\in\partial{s}}\sum_{k\in\partial{s}\setminus{i}}v_{i\rightarrow s}u_{s\rightarrow k}}{v^Tu-v^Tu_{\rightarrow s}}.
\end{equation}
With this approximation, we only need to compute the dominant eigenvalue and associated eigenvector of the nonbacktracking matrix $\textit{B}$, rather than the dominant eigenvalues of all the reduced nonbacktracking matrices, as in the MSI algorithm. So the complexity is reduced from $O(M N^{|S|})$ to $O(N^{|S|})$. The procedure of the reduced-complexity PMSI algorithm is in Table \textbf{Algorithm 2}.

\begin{algorithm}[htb]
\caption{Perturbation-based Multiple Source Identification (PMSI)}
\begin{algorithmic}[1]
\REQUIRE ~~\\
Nonbacktracking matrix $\textit{B}$ of an infected graph of $N$ nodes and $M$ edges;\\
Number of rumor sources $|S|$;
\ENSURE ~~\\
Identified rumor sources $\hat{S}$;
\FOR{$i = 1, \ldots, \binom{N}{|S|}$}
\STATE{Enumerate a set of potential source nodes as $\hat{S}_i= \{i_1,i_2,...,i_{|S|}\}$;}
\STATE{Get $u_{\rightarrow i_j}$ ($i_j\in\hat{S}_i$) from $u$ and calculate $\Delta{\lambda}_i$ according to (\ref{eqn:pmsi});}
\ENDFOR
\STATE{Declare the set $\hat{S}_i$ with the maximal $\Delta{\lambda}_i$ as rumor sources $\hat{S}$.}

\end{algorithmic}
\end{algorithm}

\section{Simulations}
\label{sec:simu}

In this section, we evaluate the performance of our proposed algorithms on different synthesized and real-world networks, including small-world networks, power grids, Facebook networks, and regular lattices.

\subsection{Single source case}
\label{subsec:single}

In single source case, we compare our algorithms with two representative algorithms, the Jordan center (JC) \cite{reference7} and the Rumor center (RC) combined with a breadth-first-search (BFS) tree heuristic \cite{reference6}. Note that for loopy networks all these algorithms are heuristic in nature.

We evaluate the performance using three metrics: (1) \emph{Accuracy}: the probability that the identified source node is the actual source. (2) \emph{One-hop accuracy}: the probability that the distance between the identified source node and the actual source is no more than one hop. (3) \emph{Average error distance}: the average number of hops between the identified source node and the actual source.

In simulating the rumor spreading process we choose $p < 0.1$ so that the infected nodes are sufficiently spread. We consider four kinds of networks: synthesized small-world networks, the western states power grid network of the United States, a fraction of the Facebook network with 4039 nodes,\footnote{Data source: http://snap.stanford.edu/data/index.html} and regular lattices. Note that these networks are all loopy, especially for the latter three kinds. We generate 500 instances of 400-node infected graphs for each network. The average diameters of infected graphs for these networks are 15.5 (small-world networks), 19.5 (power grids), 10.9 (Facebook networks) and 36.8 hops (regular lattices), respectively. Table \ref{tab1} shows the simulation results. We see that the MSI algorithm generally outperforms both JC and RC-BFS, and the PMSI algorithm also performs quite well, --- sometimes it even outperforms the original MSI algorithm. The performance advantage is the most evident for highly loopy networks, e.g., Facebook networks and regular lattices.

\begin{table}[ht]
\vspace{-0.2cm}
\caption{Simulation results in single source case}
\centering
\small
\subtable[Accuracy]{
\begin{tabular}{|c|c|c|c|c|}
\hline
\textbf{Network}&JC&RC-BFS&MSI&PMSI\\
\hline
Small-world&18.2$\%$&11.6$\%$&19.8$\%$&19.6$\%$\\
\hline
Power grids&2.6$\%$&2.6$\%$&2.6$\%$&1.4$\%$\\
\hline
Facebook&1.8$\%$&1.8$\%$&3.2$\%$&1.4$\%$\\
\hline
Regular lattices&6.8$\%$&0.6$\%$&10.4$\%$&14.4$\%$\\
\hline
\end{tabular}
}
\subtable[One-hop accuracy]{
\begin{tabular}{|c|c|c|c|c|}
\hline
\textbf{Network}&JC&RC-BFS&MSI&PMSI\\
\hline
Small-world&77.8$\%$&58.8$\%$&78.6$\%$&78.0$\%$\\
\hline
Power grids&17.2$\%$&9.6$\%$&18.2$\%$&13.8$\%$\\
\hline
Facebook&17.6$\%$&19.0$\%$&35.6$\%$&28.4$\%$\\
\hline
Regular lattices&28.2$\%$&7.0$\%$&31.6$\%$&47.4$\%$\\
\hline
\end{tabular}
}
\subtable[Average error distance]{
\begin{tabular}{|c|c|c|c|c|}
\hline
\textbf{Network}&JC&RC-BFS&MSI&PMSI\\
\hline
Small-world&1.06&1.40&1.05&1.06\\
\hline
Power grids&3.17&3.45&3.43&3.77\\
\hline
Facebook&2.37&2.35&1.96&2.13\\
\hline
Regular lattices&2.36&4.36&2.39&1.73\\
\hline
\end{tabular}
}
\label{tab1}
\end{table}

\begin{table}[ht]
\vspace{-0.2cm}
\caption{Simulation results in multiple source case}
\begin{center}
\begin{tabular}{|c|c|c|c|c|}
\hline
\textbf{Network} &$|S|$ & \textbf{\textit{Accuracy}}& \textbf{\textit{One-hop accuracy}}& $\triangle$ \\
\hline
\multirow{2}{*}{Small-world}& 2& $1.8\%$ & $26.4\%$ & 2.244  \\
\cline{2-5}
&3& $0\%$ & $22.4\%$ & 1.45 \\
\hline
\multirow{2}{*}{Facebook} & 2& $4.0\%$ & $22.2\%$ & 1.653 \\
\cline{2-5}
&3& $1.0\%$&$12.2\%$ &1.60 \\
\hline
\multicolumn{4}{l}{$\triangle$: average error distance}
\end{tabular}
\label{tab2}
\end{center}
\vspace{-0.6cm}
\end{table}

\subsection{Multiple source case}

In multiple source case, we generate 500 instances of 100-node infected graphs for small-world and Facebook networks. The sources are randomly picked among each network. The average diameters of infected graphs are 16 (small-world networks) and 8.9 hops (Facebook networks), respectively. With multiple sources, we modify the performance metrics in the following way: we associate the identified sources $\hat{S}$ with the actual sources $S$ so that the normalized total error distance between $\hat{S}$ and $S$, i.e., $\triangle = \frac{1}{|S|}\sum_{i=1}^{|S|}d(\hat{s}_i,s_i)$, where $d(\hat{s}_i,s_i)$ is the number of hops between the actual source $s_i$ and its associated identified source $\hat{s}_i$, is minimized. We then define the accuracy as the probability that $\hat{S} = S$, the one-hop accuracy as the probability that $d(\hat{s}_i,s_i) \leq 1$, $\forall i = 1, \ldots, |S|$, and the average error distance as the average of the minimum $\triangle$.

Table \ref{tab2} shows the simulation results when $|S|$ is two or three using the MSI algorithm. Although the accuracy and the one-hop accuracy drastically degrade compared with the single source case, the average error distance is usually less than two hops. Furthermore, it is interesting to notice that average error distance decreases as the number of sources increases. This may be due to that with more sources, even it is challenging to accurately identify all of them, it is likely that some of them can be accurately identified so that the average of error distances is decreased.

\section{Conclusion}
\label{sec:conclusion}

We proposed a novel heuristic source identification method for general loopy networks with multiple rumor sources, motivated by deducing and analyzing the behavior of message passing equations of the rumor spreading process, combined with some empirical observations. Numerical experiments show that for several representative kinds of general networks, the proposed method is competitive with existing methods. In future research, it is desirable to deepen our understanding of the proposed heuristic method, and to provide a solid theoretical foundation that explains its effectiveness.



\vfill\pagebreak

\bibliographystyle{IEEEbib}
\bibliography{strings}

\end{document}